\documentstyle[mprocl]{article}

\def\Lie{\hbox{\it \char'44}\!}                   

\begin{document}

\title{ENERGY-MOMENTUM (QUASI-)LOCALIZATION FOR GRAVITATING SYSTEMS}

\author{Chia-Chen CHANG, James M. NESTER, Chiang-Mei CHEN}

\address{Department of Physics, National Central University, Chungli \\
         E-mail: nester@joule.phy.ncu.edu.tw, cmchen@joule.phy.ncu.edu.tw}


\maketitle\abstracts{
Traditional approaches to energy-momentum localization led to
reference frame dependent pseudotensors.
The more modern idea is
quasilocal energy-momentum.  We take a Hamiltonian approach.
The Hamiltonian boundary term gives not only the
quasilocal values but also
boundary conditions via
the Hamiltonian variation boundary principle.
Selecting a Hamiltonian boundary term involves several choices.
We found that superpotentials can serve as Hamiltonian boundary terms,
consequently pseudotensors are actually
quasilocal and legitimate.
Various Hamiltonian boundary term quasilocal
expressions are considered including some famous pseudotensors,
 M{\o}ller's tetrad-teleparallel ``tensor'',
 Chen's covariant expressions,
 the expressions of Katz \& coworkers,
 the expression of Brown \& York,
 and some spinor expressions.
 We emphasize the need for identifying criteria for a good choice.
}

\section{Introduction}

\subsection{Energy-momentum}

Energy-momentum is a fundamental conserved quantity associated with a
symmetry of
space-time geometry.
 Noether's theorem and translation invariance leads to
 the canonical energy-momentum (EM) density tensor, $T^{\mu\nu}$, which is
 conserved: $\nabla_\nu T^{\mu\nu}=0$.
But this tensor is not unique:
one can add an arbitrary ``curl'', and it is still conserved,
 e.g., one can shift the ``zero'' of energy.
Gravity, however, responds to any real EM,
 so it removes the ambiguity in $T^{\mu\nu}$.
Thus {\em gravity absolutely identifies\/} EM.

The {\em source} of gravity is $T^{\mu\nu}$, the EM density
for matter {\em and all other} interaction fields.
But sources {\em exchange} EM
  with the gravitational field {\em locally}.
This leads to the expectation that
gravity should have its own local
energy-momentum density (GEM).

\subsection{Pseudotensors discredited}

 Attempts to identify a local
 ``{total EM density}''
 for gravitating systems, via standard techniques,
 gave only various
 non-covariant,
 reference frame dependent
{\em {pseudotensors}},
e.g., those of Einstein,\cite{Traut62} Landau-Lifshitz,\cite{LL62}
 M{\o}ller,\cite{Mol58} and Weinberg,\cite{Wein72}.  Such objects
{\em cannot} give a well defined GEM localization.
 This is in accord with the
{\em equivalence principle} (see, e.g.,\cite{MTW} p 457)
which implies that the gravitational field cannot be detected at a point.
Thus the pseudotensor approach has been {\em discredited}.

\section{Quasi-local energy-momentum}

A new idea, {\em quasi-local energy-momentum} (QEM)
 (quasilocal $\equiv$ associated with a closed 2-surface)
is now regarded as the proper fundamental notion of
energy.

There have been
many quasilocal proposals and
approaches including
 null ray geometry, \cite{Haw68}
 twistors, \cite{Pen82}
 background, \cite{Ka85,Chr85,Sor88,Pet95}
 symplectic reduction, \cite{JK87,JK90,Kij97}
 spinors,\cite{DM91} Hamilton-Jacobi,\cite{BY93}
 ``2+2'',\cite{HayS94} and
 covariant-symplectic.\cite{CNT95}
Many {\em criteria} (see, e.g.,\cite{CY88}), such as
good limits (in particular the ADM (spatial $\infty$), Bondi (null $\infty$),
weak field and flat spacetime), have have been proposed.
Positivity has also been advocated by many authors,
but it
not possible for a closed universe, and consequently,
{\em cannot} be required
in general.\cite{HayG95}
In any case, it has been noted that
an {\em infinite number} of expressions\cite{Berq92}
 satisfy all of these requirements!
Clearly there is a need for additional {\em principles} and {\em criteria}.

\section{The Hamiltonian Approach}

Energy can be defined as the
value of the Hamiltonian.
The Hamiltonian for a gravitating system
(for a finite region \& any geometric gravity theory)
 is the Noether generator of translations
 along a displacement vector field, $N^\mu$:
\begin{equation}
H(N)={{\int}_{\Sigma} N^\mu {\cal H}_\mu} +
 {{{\oint}}_{S=\partial \Sigma}{\cal B}(N)},
\end{equation}
it includes a volume (i.e., spatial hypersurface) {\em and} a
(2-dimensional) boundary term.

From Noether's theorems it follows that
${\cal H}_\mu \propto$ field equations,
consequently
$E:=H(N)$ depends only on the boundary term ${\cal B}$,
which thus gives the {\em quasilocal energy-momentum}.
However the boundary term ${\cal B}$
can be {\em modified}.
(This special case of the usual Noether current non-uniqueness is
essentially just the aforementioned ambiguity in the canonical energy-momentum
density.)\ \
 Indeed $\cal B$ {\em must} be adjusted
 to get the
correct asymptotic GR values.\cite{RT74}

\subsection{The Hamiltonian variation boundary term principle}

Fortunately, ${\cal B}$ is not arbitrary.
Its form is controlled by another principle of the formalism,
 {\em the Hamiltonian boundary variation principle}: one should
 {\em choose} the Hamiltonian boundary term ${\cal B}$
so that the boundary term in $\delta
H$ {\em vanishes},
when the desired fields are {\em fixed} on S.
There is thus a nice division:
 the Hamiltonian density ${\cal H}_\mu$
determines the evolution and constraint equations,
 the boundary term ${\cal B}$
determines the boundary conditions and the QEM.

For Einstein's GR theory,
(succinctly in terms of differential forms)
 with the connection one-form $\omega^\alpha{}_\gamma$,
the curvature 2-form  $\Omega^\alpha{}_\gamma:=d\omega^\alpha{}_\gamma
+\omega^\alpha{}_\mu\wedge\omega^\mu{}_\gamma$,
and  $\eta^{\alpha\gamma
\cdots}:=*(\vartheta^\alpha\wedge\vartheta^\gamma\cdots)$,
 where $\vartheta^\alpha$ is the coframe one-form, we
 space-time split the Hilbert Lagrangian:
\begin{equation}
{\cal L} = dt \wedge i_N {\cal L} =
dt \wedge i_N (\Omega^\alpha{}_\gamma \wedge\eta_\alpha{}^\gamma)
= dt \wedge [\Lie_N \omega^\alpha{}_\gamma \wedge
\eta_\alpha{}^\gamma - {\cal H}(N)],
\end{equation}
to identify the {\em Covariant Hamiltonian} as
\begin{equation}
{\cal H}(N) = -N^\mu \Omega^\alpha{}_\gamma \wedge
\eta_\alpha{}^\gamma{}_\mu
- i_N \omega^\alpha{}_\gamma\, D \eta_\alpha{}^\gamma
+ d (i_N \omega^\alpha{}_\gamma \, \eta_\alpha{}^\gamma)\, .
\end{equation}
The total derivative term here integrates,
 via the generalized Stokes theorem,
to a boundary term --- which needs adjustment.

\subsection{Covariant Boundary Terms}

Although there are an infinite number of possible boundary terms,
for {\em each} dynamical field C.M. Chen found,\cite{CNT95,CN99}
using {\em sympletic} techniques,\cite{KT79}
 that
there are {\em only two} {\em covariant choices} for $\cal B$,
depending upon what is held fixed on the boundary:
\begin{eqnarray}
{\cal B}_\vartheta  &=&
  \Delta \omega^\alpha{}_\gamma \wedge i_N \eta_\alpha{}^\gamma +
i_N {\buildrel \scriptstyle \circ \over
 \omega}{}^\alpha{}_\gamma \, \Delta \eta_\alpha{}^\gamma  \,
\label{Btheta},\\
 {\cal B}_\omega  &=&
  \Delta \omega^\alpha{}_\gamma \wedge i_N  {\buildrel
 \scriptstyle \circ \over \eta}{}_\alpha{}^\gamma +
 i_N \omega^\alpha{}_\gamma\, \Delta \eta_\alpha{}^\gamma \, ,
\end{eqnarray}
where for any quantity $\Delta\alpha:=\alpha-{\buildrel \scriptstyle \circ
\over \alpha}$, with
${\buildrel \scriptstyle \circ \over \alpha}$
determining a fixed {\em reference
configuration}.
For these choices of Hamiltonian boundary term, the boundary term in $\delta H$
is  a projection of a 4-covariant expression along the displacement
vector field, respectively, \begin{equation}
di_N(-\Delta\omega^{\alpha}{}_\gamma\wedge\delta \eta_\alpha{}^{\gamma}),
\qquad
di_N(\delta\omega^{\alpha}{}_\gamma\wedge\Delta \eta_\alpha{}^{\gamma})
,
\end{equation}
representing
a {\em Dirichlet} or {\em Neumann} type ``control mode''.

In passing, we call attention to the important {\em gauge term} $i_N\omega$.
It plays a role in generating the correct dynamic evolution of the
frame along with the proper variational boundary condition.   However it also
adds
an unphysical, noncovariant ``rotation of the reference frame''  contribution
to the quasilocal energy-momentum.  This contribution can be isolated by
inserting the identity
 \begin{equation}
(i_N \omega^\alpha{}_\gamma) \vartheta^\gamma \equiv
 D N^\alpha
 - \Lie_N \vartheta^\alpha.
\end{equation}
It can then be removed simply by dropping the terms
proportional to $\Lie_N\vartheta$.

Our ``covariant'' quasilocal expressions yield the
 correct total ADM and Bondi values.\cite{HN93,HN96}
Also, they have widely accepted limiting forms; in fact for many
practical cases our
quasilocal energy (likewise momentum) essentially reduces to the well known
quasilocal expression of Brown \& York,\cite{BY93}
\begin{equation}
E=\oint (K-K_0),
\end{equation}
although in general there are some important differences: in
particular (i) our expressions are {\em 4-covariant},
(ii) they allow for general reference configurations,
(iii)  and more general displacements,
(iv) they include a M{\o}ller-Komar like term,
and (v) our boundary need not be orthogonal to $\Sigma$.

\subsection{Applications}

As an example consider a static spherically symmetric star. In the
exterior, where the geometry has the Schwarzschild form, for the
quasilocal energy within a sphere (using
Dirichlet boundary conditions, a Minkowski reference configuration and
the Minkowski timelike Killing vector for the displacement) we find
\begin{equation}
E=r(1-\sqrt{1-2M/r})=M(1+M/2R)
\end{equation}
in terms of the Schwarzschild $r$ (with necessarily $r>2M$) and isotropic $R$,
respectively.
Note that this is a {\em decreasing} function which approaches $M$
asymptotically.
Hence, for a region between two spherical boundaries in the
exterior, the quasilocal energy is {\em negative}.  Extending this to
a black hole, using an inner boundary on the horizon and outer
boundary at infinity, (according to this definition) the outer
integral gives $M$ and the inner gives $2M$ so that the total value of
the quasilocal energy within the intermediate region is $-M$.

A related application of this formalism is to black hole
thermodynamics.  We considered the appropriate Hamiltonian for the
region between an inner boundary on the horizon and a boundary at
$\infty$ and obtained the first law.  The outer integral gave the
total energy and work terms, the integral over the horizon gives the
expression for the entropy.\cite{CNT95,CN99}

In an orthonormal Cartesian frame
our Hamiltonian boundary term becomes
\begin{equation}
{\cal B}_T:=\omega^{\alpha\gamma}\wedge i_N\eta_{\alpha\gamma},
\end{equation}
which yields M{\o}ller's tetrad-teleparallel energy-momentum
``tensor''.\cite{Mol61}

In a holonomic (coordinate) basis, our covariant boundary expressions
correspond to
controlling
$\pi^{\alpha\sigma}:=\sqrt{-g}g^{\alpha\sigma}$ (Dirichlet), or
$\Gamma^\alpha_{\gamma\lambda}\sim \partial g$ (Neumann),
respectively, are  \begin{eqnarray}
\!\!\!\!
{{\cal B}}_{g}
  &=& N^\mu \pi^{\gamma\sigma}
\Delta\Gamma^\alpha_{\gamma\lambda}\,
 \delta^{\tau\rho\lambda}_{\alpha\sigma\mu}+
  {\buildrel \scriptstyle \circ \over  D}{}_\gamma N^\alpha \Delta
(\pi^{\gamma\sigma})\delta^{\tau\rho}_{\alpha\sigma},  \\
\!\!\!\!
{{\cal B}}_{\Gamma}&=&
 N^\mu
{\buildrel \scriptstyle \circ \over  \pi}{}^{\gamma\sigma}
\Delta\Gamma^\alpha_{\gamma\lambda}
\delta^{\tau\rho\lambda}_{\alpha\sigma\mu}+   D_\gamma N^\alpha
\Delta(\pi^{\gamma\sigma})\delta^{\tau\rho}_{\alpha\sigma}.
\end{eqnarray}
Our {\em Dirichlet} mode matches the recent expression of Katz, Bi{\v
c}\'ak \& Lynden-Bell,\cite{KBLB97} and Katz \& Lerer,\cite{KL97}
which they derived via a Noether argument applied to gravity with a
fixed global
background (whereas we require our reference configuration only on the
boundary).  Their work includes a discussion of applications of this
expression to cosmology and Mach's principle.  By modifying their
construction we can likewise obtain our alternative {\em Neumann}
control mode.

\subsection{Choices}

Selecting a Hamiltonian boundary term involves many choices:

\begin{description}
\item[(1)] the {\em representation} or choice of {\em dynamic
variables}: metric, orthonormal frame, connection, spinors \dots;

\item[(2)] the {\em control mode} or {\em boundary conditions}: e.g.,
covariant Dirichlet/Neumann;

\item[(3)] the {\em reference configuration}: e.g.,
Minkowski, (anti-)de Sitter, FRW cosmology, Schwarzschild;

\begin{itemize}
\item The physical meaning is that all of the quasilocal quantities {\em
vanish}  when the field has the reference values, so it
fixes the zero of energy, etc.

\item This could be determined either
geometrically (by embedding and matching conditions) or
dynamically (by an equation).

\end{itemize}

\item[(4)] the {\em displacement vector field} $N$:
 which timelike displacement gives the energy?
 which spatial displacement gives the momentum?
A good choice is to use a Killing vector of the reference
geometry.\cite{CN99}

\end{description}

As a consequence of such choices there are various kinds of energy,
each associated with specific boundary conditions.  The physical
situation may be compared with thermodynamics (where we have enthalpy,
Gibbs, Helmholtz, \dots) or electrostatic energy in a region with
fixed surface potential vs.\ fixed surface charge density.

\section{Pseudotensors Rehabilitated}

Consider the pseudotensor idea in detail.{\cite{CNC99}}
First select a {superpotential} ${H_\mu{}^{\nu\lambda}}\equiv
H_\mu{}^{[\nu\lambda]}$, and then use it to split the Einstein
tensor by defining
the GEM pseudotensor:
\begin{equation}
\kappa \sqrt{-g} N^\mu {t_\mu{}^\nu}:=
-N^\mu\sqrt{-g}G_\mu{}^\nu+{1\over2} \partial_\lambda( N^\mu
{H_\mu^{\nu\lambda}}).
\end{equation}
Then Einstein's equation, $G_\mu{}^\nu=\kappa T_\mu{}^\nu$,
(with constant $N^\mu$)
takes a form with the
{{\em total} effective  EM pseudotensor} as its source:
\begin{equation}
\partial_\lambda {H_\mu^{\nu\lambda}}=
2\kappa (-g)^{1\over2}{{\cal T}_\mu{}^\nu}:=
2\kappa(-g)^{1\over2}( {t_\mu{}^\nu} + T_\mu{}^\nu).
\end{equation}
Consequently
$\partial_\nu (-g)^{1/2} {{\cal T}_\mu{}^\nu}\equiv0$,
which thus integrates to a conserved energy-momentum:
\begin{equation}
N^\mu P_\mu:=\int  N^\mu {{\cal T}_\mu{}^\nu}
(-g)^{1/2}(d^3x)_\nu,
\end{equation}
whereas
$\nabla_\nu T_\mu{}^\nu=\partial_\nu
T_\mu{}^\nu-\Gamma^\lambda{}_{\nu\mu}T_\lambda{}^\nu+\Gamma^\nu{}_{\lambda\nu}T
_\mu{}^\lambda=0$
{\em does not} --- unless $\Gamma=0$ (flat space).

Minor variations of this procedure use
 ${H^{\mu\nu\lambda}}\equiv H^{\mu[\nu\lambda]}$,
 and even further
 ${H^{\mu\nu\alpha}}:=\partial_\gamma {H^{\mu\alpha\nu\gamma}}$,
with
 ${H^{\mu\alpha\nu\gamma}}\equiv
H^{\nu\gamma\mu\alpha}\equiv H^{[\mu\alpha][\nu\gamma]}$ and
$H^{\mu[\alpha\nu\gamma]}\equiv0$.
The latter guarantees a symmetric pseudotensor and thus a simple
definition of angular momentum.
A special case,\cite{Gold58}
\begin{equation}
{H^{\mu\alpha\nu\gamma}}(h):={h}^{\mu\nu}{h}^{\alpha\gamma} -
{h}^{\alpha\nu}{h}^{\mu\gamma},
\end{equation}
 generalizes to
\begin{equation}
{H}({h_1},{h_2})
:={H}({h_1+h_2})-{H}({h_1})-{H}({h_2}).
\end{equation}
For suitable choices of ${H}$, ${h}$, ${h_1}, {h_2}$, these
expressions include {\em all} of the famous pseudotensors.
Formally,
including the displacement vector field,
 and making  adjustments like
 $N^\mu H_\mu{}^{\nu\lambda}\longrightarrow N_\mu H^\mu{}^{\nu\lambda}$
allows us to cover all these particular forms.

We note that the superpotentials can serve as
 Hamiltonian boundary terms and consequently the associated pseudotensors are
fundamentally {\em quasilocal}:
\begin{eqnarray}
-P(N)
&:=&-\int_{\Sigma} N^\mu {{\cal T}_\mu{}^\nu} \sqrt{-g}(d^3x)_\nu
\equiv\int_{\Sigma} N^\mu \sqrt{-g}(-t_\mu{}^\nu-T_\mu{}^\nu)
(d^3x)_\nu \nonumber\\
&\equiv&
\int_{\Sigma} \bigl[ N^\mu\sqrt{-g}({1\over\kappa}
G_\mu{}^\nu-T_\mu{}^\nu)
- {1\over2\kappa}\partial_\lambda (N^\mu {H_\mu{}^{\nu\lambda}})\bigr]
(d^3x)_\nu\nonumber\\
&\equiv&\int_{\Sigma} N^\mu {\cal H}_\mu +
 \oint_{S=\partial \Sigma}{\cal B}(N)\equiv H(N),
\end{eqnarray}
here
${\cal H}_\mu$ is the covariant form of the ADM Hamiltonian
and
\begin{equation}
{\cal B}(N)=-N^\mu (1/2\kappa)
{H_\mu{}^{\nu\lambda}}(1/2) (d^2x)_{\nu\lambda}.
\end{equation}

In all cases, to understand the {\em physical meaning} of the associated
quasilocalization we
calculate the Hamiltonian variation:
\begin{equation}
\delta H(N)=\int_\Sigma \hbox{field eqn.\ terms}
 +
 \oint_{\partial\Sigma} \hbox{ham.\ var.\
bound.\ term}.
\end{equation}
The key is the
 {\em Hamiltonian variation boundary term:}
\begin{equation}
-{1\over4\kappa}\left[\delta\Gamma^\alpha{}_{\gamma\lambda}N^\mu
\pi^{\gamma\sigma}\delta^{\tau\rho\lambda}_{\alpha\sigma\mu}
+\delta(N^\mu {H_\mu{}^{\tau\rho})}\right]dS_{\tau\rho}.
\end{equation}
It shows what must be held fixed on the boundary.

Let us now consider some specific examples, in each case
taking our
 reference configuration as Minkowski space and using
 Cartesian coordinates.
The
{\em Einstein pseudotensor} follows from the {\em Freud}
 superpotential,\cite{Freud}
 \begin{equation}
{H_\lambda{}^{\mu\nu}}
=\pi^{\gamma\sigma}\Gamma^\alpha{}_{\gamma\rho}
\delta^{\mu\nu\rho}_{\alpha\sigma\lambda}\equiv
{g_{\lambda\alpha}\over\sqrt{-g}}\partial_\gamma
H^{\mu\alpha\nu\gamma},
\end{equation}
where
$H^{\mu\alpha\nu\gamma}:=
\pi^{\mu\nu}\pi^{\alpha\gamma}-\pi^{\mu\alpha}\pi^{\nu\gamma}.$
We find
\begin{equation}
\delta H=\hbox{field eqn.\ terms}
+
\oint \delta(N^\mu\pi{}^{\gamma\sigma})
\Gamma^\alpha{}_{\gamma\lambda}
\delta^{\tau\rho\lambda}_{\alpha\sigma\mu} (d^2x)_{\tau\rho}.
\end{equation}
Hence we should use constant
$N^\alpha$ and control
$\pi^{\alpha\gamma}:=\sqrt{-g} g^{\alpha\gamma}$ on the
boundary.
For the
{\em Landau-Lifshitz} pseudotensor:
 $L^{\mu\nu}:=
\partial_\gamma \partial_\alpha H^{\mu\nu\alpha\gamma}
\equiv
\partial_\alpha(
\pi^{\mu\sigma} H_\sigma{}^{\alpha\gamma}),$
by a similar calculation we find that we should take
 $N^\alpha=\pi^{\alpha\gamma}N^0_\gamma$
                   for fixed $N^0_\gamma$,
and control $g^{\mu[\alpha}g^{\gamma]\nu}$.
For the {\em M{\o}ller} pseudotensor,\cite{Mol58} which has the
superpotential
 \begin{equation}
H_\lambda{}^{\alpha\sigma} = 2
 \pi^{\nu[\sigma}\Gamma^{\alpha]}{}_{\nu\lambda}=
2\pi^{\nu\sigma}g^{\alpha\mu} \partial_{[\nu}g_{\mu]\lambda},
\end{equation}
we must control the connection $\Gamma \sim\partial g$.

A similar analysis can be applied to other expressions old and new ---
e.g., for the recently proposed ``symmetric'' expression of
Petrov \& Katz,\cite{PK99}
\begin{equation}
{\cal B}_{KP}:=\Delta\pi^{\sigma[\tau}
{\buildrel \scriptstyle \circ \over D}{}_\sigma N^{\rho]}+
N_\lambda{\buildrel \scriptstyle \circ \over D}{}_\gamma\left(
\Delta\pi^{\alpha[\lambda}{\buildrel \scriptstyle \circ \over
g}{}^{\gamma]\sigma}\right)\delta^{\tau\rho}_{\alpha\sigma},
\end{equation}
we expect to get a mixed boundary condition.
\bigskip

\section{Spinor Formulations}

Using a spinor parameterization
$N^\mu ={\overline\psi}\gamma^\mu\psi$,
 and adjusting the boundary
term via a certain {\em spinor-curvature identity\/}\cite{NTZ94}
yields the Hamiltonian
associated with the {\em Witten} positive energy
proof,\cite{N84,CNT95} \begin{eqnarray}
{\cal H}_{w}(\psi)
 &:=&
 4 D{\overline\psi}\wedge \gamma_5 \vartheta \wedge D \psi
- i_N\omega^{\alpha\gamma}\,  D \eta_{\alpha\gamma}\nonumber\\
&\equiv &
 - N^\mu \Omega^{\alpha\gamma}\wedge\eta_{\alpha\gamma\mu}
-i_N\omega^{\alpha\gamma}\,  D \eta_{\alpha\gamma}
 + d{\cal B}_{w}\, ,
\end{eqnarray}
where $\vartheta:=\vartheta^\mu\gamma_\mu$ and
\begin{equation}
{\cal B}_{w}:=2( {\overline \psi}
\gamma_5 \vartheta \wedge D \psi
   + D {\overline \psi} \wedge\gamma_5 \vartheta \psi).
\end{equation}
 Several similar quasilocal
 boundary expressions have
 been
 investigated, e.g.,\cite{DM91,Sz94}.
The {\em beauty} of these spinor formulations is that
 they do not need
an {\em explicit} reference configuration.
We introduce a
reference configuration here
 only to compare with our other
formulation.
With
$D = {\buildrel \scriptstyle \circ \over D} + \Delta \omega$, get
\begin{equation}
{\cal B}_{w}(\psi)=
 {\overline \psi} \gamma^\mu \psi
\Delta  \omega^{\alpha\gamma} \wedge \eta_{\alpha\gamma\mu}
+
 2({\buildrel \scriptstyle \circ \over D} {\overline \psi} \wedge
\gamma_5  \vartheta \psi + {\overline \psi} \gamma_5 \vartheta \wedge
{\buildrel \scriptstyle \circ \over D} \psi).
\end{equation}
The first term is the same as the main term in eq (\ref{Btheta}).

Another class of spinor quasilocal boundary expressions
 has
been identified.
  Via other {\em spinor-curvature\/} identities\cite{NTZ94}
{\em quad\-ratic spinor\/}
 Lagrangians (QSL) for GR
have been found.\cite{NT95,TN99}
 The
Einstein-Hilbert scalar
curvature Lagrangian equals (up to an exact differential) the QSL
\begin{equation}
{\cal L}_{qs}:=2D{\overline \Psi}\gamma_5D\Psi
\equiv R*1 + d(D{\overline \Psi}\gamma_5\Psi
+{\overline \Psi}\gamma_5D\Psi)
\end{equation}
where $\Psi=\vartheta\psi$ is a spinor one-form field.
The corresponding first order Lagrangian:
\begin{equation}
{\cal L}_{\Psi}:=D{\overline\Psi} P+{\overline P}D\Psi+
{\textstyle{1\over2}}{\overline P}\gamma_5 P\, ,
\end{equation}
yields the first order equations
\begin{equation}
{\overline P}=2D{\overline \Psi}\gamma_5,\quad
P=2\gamma_5D\Psi, \qquad DP=0=D{\overline P},
\end{equation}
and can be used to construct
the covariant Hamiltonian 3-form:
\begin{eqnarray}
{\cal H}_{\Psi}
&:=&{\overline P}\Lie_N\Psi+\Lie_N{\overline \Psi} P
-i_N{\cal L}_{\Psi}\nonumber\\
& \equiv&
-i_N({\textstyle{1\over2}}{\overline P}\gamma_5 P)
-\left[i_N{\overline \Psi}DP+D{\overline \Psi}i_N P
+{\overline\Psi}i_N\omega P
-
 d(i_N{\overline \Psi}P)+ \hbox{c.c.}\right] .
\end{eqnarray}
The associated Hamiltonian boundary term
\begin{equation}
{\cal B}(N)=i_N{\overline\Psi}P + {\overline P}i_N\Psi\, ,
\end{equation}
 and
the Hamiltonian boundary variation symplectic
structure
\begin{equation}
\delta{\cal H}_\Psi(N)\simeq di_N
 (\delta{\overline \Psi}\wedge P + {\overline P} \wedge \delta \Psi)\, ,
\end{equation}
are simple.
Both are independent of any reference configuration.
But we can introduce a
reference configuration
 replacing the boundary term by
one of
\begin{eqnarray}
{\cal B}_{\Psi}&:=& i_N{\overline \Psi} \Delta P
+\Delta {\overline \Psi} i_N {\buildrel \scriptstyle \circ \over P}
+ \hbox{c.c.}\ , \\
{\cal B}_{P}&:=&
 i_N{\buildrel \scriptstyle \circ \over{\overline \Psi}}
\Delta P +\Delta {\overline \Psi} i_N  P +\hbox{c.c.} .
\end{eqnarray}
Then the variation of the Hamiltonian will contain one of the boundary
terms
\begin{equation}
i_N\left(\Delta {\overline P}\wedge\delta \Psi + \hbox{c.c.}\right)\ ,
 \quad \hbox{or} \quad
i_N\left(-\delta{\overline P}\wedge\Delta\Psi + \hbox{c.c.}\right) \ .
\end{equation}

Several remarks are in order.
\begin{description}
\item[(1)]
Note that there are
no explicit connection
terms in these spinor boundary terms. However
they are present implicitly via $P=2\gamma_5D\Psi$.
\item[(2)] Only if we include an explicit reference configuration do we get
$i_N\omega\sim DN$ M{\o}ller-Komar type terms.  Such terms
have often been overlooked in earlier investigations.
However we have found
that they
play a key role both in
certain angular momentum calculations\cite{HN93,HN96} and in
black hole
thermodynamics.\cite{CNT95,CN99}
\item[(3)]
In addition to
fixing the frame, the spinor field should be held fixed on the
boundary.
  The physical meaning of fixing the spinor field is
not yet so clear.
\end{description}

\section{Concluding Discussion}

We have noted that EM is both related to a symmetry under
space-time displacements and is also the source of gravity.
 Sources interact with gravity and thus should exchange
EM with the gravitational field yet there is no proper local
density for gravitational or total EM; in the
light of the equivalence principle it is not surprizing
that traditional methods lead only to reference frame dependent
{\em pseudotensors}.

It is now widely believed that EM is really {\em
quasilocal}. Identifying energy with the value of the Hamiltonian, we
note that boundary terms in the Hamiltonian for a finite region
(i) give the quasilocal
EM, and (ii) via the {\em Hamiltonian boundary variation
principle}, reflect the ``control mode'', i.e., the  boundary
conditions.

There are many  {\em choices} involved in selecting a Hamiltonian
boundary term, e.g.,   variables, control mode,
 displacement, reference configuration.
 For GR we found that there are only two {\em covariant} control
modes.  Schwarzschild and black-hole thermodynamics applications
were noted as well as a connection with the work of Brown \& York
and Katz and coworkers.

One class of Hamiltonian boundary term choices is to use a
superpotential.  In this way we show that the pseudotensors are
actually quasilocal and legitimate expressions for EM.  Each
pseudotensor is then associated with the value of the Hamiltonian which
evolves the variables subject to certain boundary conditions.  Several
examples of pseudotensors and quasilocal expressions were considered.

We also discusssed how our formalism deals with a couple of
spinor quasilocal EM formulations.

In conclusion we stress the importance of identifying appropriate
quasilocal energy-momentum criteria.
In addition to good asymptotic and weak field/empty space limits
(and realizing that quasilocal energy need not be positive) we see a
need to
relate the choice of boundary conditions to the physics.  For
electrodynamics (or thermodynamics) we have a good idea of the
physical significance of our boundary conditions, but not for gravity.
What does it mean to hold the metric (connection) fixed on the
boundary (i.e., give it a prescribed time dependence)?  What boundary
condition corresponds to ``thermal insulation'', i.e., no flow of
energy-momentum through the boundary?
 Suppose that observers measured
the metric and connection coefficients on the walls, floor and
ceiling of this room, what component values in what reference frame
would indicate that it was devoid of energy-momentum?
Answering such questions should
contribute much to our understanding of energy-momentum and its
localization.

\section*{Acknowledgments}
We are very grateful to J.P. Hsu, L. Liu and K. Nandi for useful comments.
This work was supported by the National Science Council
under grants NSC 89-2112-M-008-020 \& 89-2112-M-008-016.

\end{document}